\documentclass[onecolumn,12pt]{article}  
\usepackage{amssymb}
\usepackage{graphicx}
\usepackage{dcolumn}
\usepackage{bm}
\usepackage{epsfig}

\RequirePackage{graphicx}

\newcommand{\co}[1]{#1}

\begin{document}

\title{Interacting Ricci-like holographic dark energy}

\author{Fabiola Ar\'evalo$^1$\footnote{fabiola.arevalo@ufrontera.cl}, Paulo Cifuentes$^1$\footnote{p.cifuentes@ufromail.cl},  Samuel Lepe$^2$\footnote{slepe@ucv.cl} and Francisco Pe\~{n}a$^1$\footnote{francisco.pena@ufrontera.cl}\\ \small{
$^1$ Departamento de Ciencias F\'isicas, Facultad de Ingenier\'ia y Ciencias, }\\
\small{Universidad de La Frontera, Temuco, Casilla 54-D, Chile }\\
\small{$^2$ Instituto de F\'isica, Pontificia Universidad Cat\'olica de Valpara\'iso,}\\ \small{Casilla 4950, Valpara\'iso, Chile.}}

\date{\today }
\maketitle
\begin{abstract}
In a flat Friedmann-Lema\^{i}tre-Robertson-Walker background, a scheme of dark
matter-dark energy interaction is studied considering a holographic
Ricci-like model for the dark energy. Without giving a priori some
specific model for the interaction function, we show that this function
can experience a change of sign during the cosmic evolution. The
parameters involved in the holographic model are adjusted with Supernova data and we obtained results compatible with the observable universe.
\end{abstract}

\maketitle
\section{Introduction}
\label{intro}
The observational data from anisotropic measures of the cosmic microwave background
\cite{Komatsu:2010fb} and observations of type Ia Supernova  \cite{Riess:1998cb,Perlmutter:1998np}, among other, shows that the universe has entered a phase of late accelerated expansion. The most accepted interpretation in the context of Einstein's General Relativity is that this phase is driven by a kind of unknown component called dark energy \cite{Copeland:2006wr,Frieman:2008sn}, which constitutes around $73\%$ of the total energy density in the universe today.
This component is usually described by a fluid with a negative pressure and various theoretical models have been proposed, among others, scalar field models called quintessence 
\cite{Ratra:1987rm,Dai:2009mq,Zlatev:1998tr}, tachyon fields 
\cite{Sen:2002in,Abramo:2003cp}, quintom (schemes that cross the constant barrier of the cosmological constant $\omega_{_{DE}}$=-1) \cite{Elizalde:2004mq,Feng:2004ad},
phantom fields (schemes that leads to future singularities in the evolution) \cite{Caldwell:1999ew,Nojiri:2003vn} and holographic dark energy 
\cite{Hu:2006ar}, which is based on the application of the holographic principle to cosmology. 
According to \cite{Cohen:1998zx} the energy contained in a region of size $L$ must not exceed the mass of a black hole of the same size, which means, in terms of energy density, $\rho\leq
L^{-2}$. If in a cosmological context we saturate this inequality by doing $H=L^{-1}$, where $H$ is the Hubble parameter, as $\rho\sim H^{2}$, we obtain a model of holographic nature for the dark energy density. Based on this idea, M. Li \cite{Li:2010cj,Li:2004rb} proposed the model $\rho=3\tilde{c}^{2}{H^{2}}$, where $\rho$ is the dark energy density and $\tilde{c}^2$ is a positive constant that fulfills $\tilde{c}^{2}<1$ \cite{Zhang:2013mca}. \co{In \cite{Nojiri:2005pu} there is an expression for the pressure as a function of $p=p \left(H^2,\dot H \right)$ 
 and there is a generalization of holographic dark energy as a function of the particle horizon and the future horizon, i.e. $L_{\Lambda}=L_{\Lambda} \left( L_p,L_f,t_s,H,l\right)$ }. 
Later on the model $\rho\sim R$ was proposed, where
$R=6\left(  {2{H^{2}}+\dot{H}+k/{a^{2}}}\right)  $ is the Ricci scalar \cite{Gao:2007ep}. %X
In the reference \cite{Granda:2008dk}, inspired by the curvature scalar and for the spatially flat section, the model called Ricci-like was then proposed, with  $\rho=3\left(  {\alpha{H^{2}}+\beta \dot{H}%
}\right)  $, where $\alpha$ and $\beta$ are constant. Moreover, this holographic Ricci-like dark energy density is also proposed in \cite{Granda:2011zm} as an expansion on powers of its arguments corresponding to a cut-off  given by a general function that depends on the Hubble parameters and its first derivative.\\ 
In the context of non-interacting cosmic fluids, where dark matter and energy are conserved independently, both constants are positive and less than one according to the supernovae comparison performed by \cite{Lepe:2010vh}. Interacting models based on this generalization have been studied by \cite{Chimento:2011pk} for the modified holographic Ricci Dark Energy (MHRDE), although this work was focused on providing an interaction function rather than obtaining it.

The present work was developed in the framework of a flat Friedmann-Lemaitre-Robertson-Walker (FLRW) cosmology with two interacting fluids, with one of these being a dark energy component (Ricci-like) and the other component, a pressureless dark matter component.
The interaction form will not be given a priori as an Ansatz, but rather as a function to be determined.
The possibility that this function crosses the non-interacting line makes studying the interacting model interesting, given that the phenomenological standard choices do not permit a transition. The sign of the phenomenological interacting term, $Q$, determines the direction of the energy flux. If $Q$ is positive, there is a decay of dark energy into dark matter and if $Q$ is negative, there is transference of dark matter into dark energy. Observational evidence has been found of the interaction between dark energy and dark matter from the Abell Cluster
A586 \cite{Bertolami:2007zm}. Observations by \cite{Pereira:2008at,Costa:2009mv,Guo:2007zk} indicate the possibility of the decay of dark matter into dark energy and more recently \cite{Costa:2013sva} analyze linear cosmological interaction with Planck data and conclude that the interaction is compatible with the latest and more precise observations. It is noteworthy to mention that they obtain slightly negative interaction in some cases. 
In \cite{Xu:2013jma} the authors contrast dark sector interactions with three different observables show a relation between the sign and amplitude of the interaction with the amplitude of the kSZ effect. They also mention that future experiments might provide the first detection of interaction in the dark sector. \\
We will discuss the characteristic of such interaction (described through a function $Q$), focusing on the changes it undergoes in the presence of the term ${\dot{H}}$.
 We will study the interaction between dark energy and dark matter in the context of holographic Ricci-like dark energy. 
  Different Ansatzes for the variable state parameters will be implemented to close the system, such that it can be integrated analytically. This allows the energy densities to be obtained, and the resulting ranges where the weak energy condition (positive energy densities) is respected. \\ 
Finally, we will adjust the parameters $\alpha$ and $\beta$ with type Ia Supernova data and thus describe the function $Q$. \co{Such function is a result of the holography and its sign during the evolution of the system is related to the parameters value, not a prior. 
}We will also discuss different models of late evolution that result from our approach. \\
This article is organized as follows, Section $2$ describes the scheme of interacting fluids with holographic Ricci-like dark energy and Section $3$ develops some models in detail using observational constraints with Type Ia supernovae. Finally, Section $4$ is devoted to the discussion and final remarks.

\section{Interacting fluids and Q-function} \label{dos}
We begin our analysis considering an interacting flat model with $\rho_{_{DE}}$, the dark energy component and the pressure $p_{_{DE}}$ and $\rho_{_{DM}}$, the pressureless dark matter component. The field equations, considering $8 \pi G$=$c$=$1$,  are 
\begin{eqnarray}
3H^2&=&\rho_{_{DE}}+\rho_{_{DM}} ,\label{00}\\
 \dot H + {H^2} &=&  - \ \frac{ 1}{6}\left({\rho _{_{DM}}} + {\rho _{_{DE}}} + 3p_{_{DE}}+3p_{_{DM}}\right) , \label{yy}
\end{eqnarray}
where $H=\dot a/ a$ is the Hubble parameter and $a$ is the scale factor. 
Assuming that in our model, dark energy exchanges energy with dark matter  through a phenomenological coupling function $Q$, the conservation equation for each fluid are given as
\begin{eqnarray}
\dot{\rho}_{_{DE}}+3H\left(\rho_{_{DE}}+p_{_{DE}}\right)&=&-Q , \label{DE}\\
\dot{\rho}_{_{DM}}+3H \left(\rho_{_{DM}}+p_{_{DM}} \right)&=&Q .\label{DM}
\end{eqnarray}
There have been several models considered in the literature for $Q$ as a function of the energy densities and the Hubble parameter \cite{Chen:2011cy}, usually considering a specific form that allows the equations (\ref{DE}) and (\ref{DM}) to be integrated analytically \cite{Chimento:2009hj}. In the present study, we do not use a specific model for this coupling term, but rather obtain it after implementing, in the field equations, different Ansatzes, either for the EoS parameter or the cosmic coincidence parameter $r=\frac{\rho_{DM}}{\rho_{DE}}$ \footnote{ As we will see, both alternatives will prove to be equivalent.}. From (\ref{DE}) and (\ref{DM}) the total energy density equation is conserved
\begin{eqnarray}
\dot{\rho}_{_{DE}}+\dot{\rho}_{_{DM}}+3H(\rho_{_{DE}}+\rho_{_{DM}}+p_{_{DM}}+p_{_{DE}})=0
. \label{DE+DM}
\end{eqnarray}

For the dark energy density we consider the Ricci-like model given by
\begin{eqnarray}
\rho_{_{DE}}=3(\alpha H^2+\beta\dot{H}) , \label{HL}
\end{eqnarray}
where $\alpha$ y $\beta$ are both positive constants that  will be fit using observational data.

\co{This Ansatz has an implicit equivalence with others models for the dark energy pressure described in \cite{Capozziello:2005pa}, which can be clearly seen obtaining $\dot{H}$ from (\ref{HL}) and replacing it in (\ref{yy}) for a dust dark matter configuration, as
\begin{eqnarray}
p_{_{DE}}=-\frac{2}{3 \beta} {\rho}_{_{DE}}+\frac{ H^2}{\beta}\left( 2 \alpha-3\beta \right), \label{pDE}
\end{eqnarray}
which is equivalent to the initial pressure described in the Increased Matter model for $\omega_f=-\frac{2}{3 \beta} $ and $\omega_H=\frac{1}{\beta}\left( 2 \alpha-3\beta \right)$. This can also be seen as a Quadratic Equation of State $F({\rho}_{_{DE}},p_{_{DE}},H)=0$ resolving (\ref{pDE}) to be zero. All of these pressures emerge as a particular case of a generalized fluid that uses a function $f(t)$ for the Hubble parameter in order to study the pressure as a function of $(\rho,f)$, which is equivalent to consider the Ricci-like dark energy model. \\
The expression (\ref{pDE}) is not politropic but resembles inhomogeneous fluid cosmology, where the pressure has dependence not only in the dark energy density but also on the Hubble parameter. In \cite{Bamba:2012cp} the authors review this model considering an expression for the dark energy pressure as a function of $({\rho},H, \dot H)$ which for our case can be seen rewritting the pressure in eq (\ref{pDE}). \\
We shall consider a pressureless dark matter content $p_{_{DM}}=0$ and $\omega_{_{DE}}=p_{_{DE}}/\rho_{_{DE}}$ as the barotropic equation of state (EoS) for the dark energy, which varies with time.} \\
By taking the time derivative of (\ref{00}) with respect to time, we obtain
\begin{eqnarray}
6H\dot{H}=\dot{\rho}_{_{DE}}+\dot{\rho}_{_{DM}} , \label{dv00}
\end{eqnarray}
and using (\ref{00}), (\ref{HL}) and (\ref{dv00}) in (\ref{DE+DM}), a differential equation is obtained for the Hubble parameter in terms of the parameter $\omega_{_{DE}}$,
\begin{eqnarray}
\dot{H}=-3\left( \frac{1+\alpha \ \omega_{_{DE}}}{2+3 \ \beta \  \omega_{_{DE}}}\right) H^2 , \label{1Hpt}
\end{eqnarray}
an equation that cannot be integrated, unless we have information about the dark energy variable parameter $\omega_{_{DE}}$. In this case, the holographic Ricci-like dark energy can be written in the form
\begin{eqnarray}
\rho_{_{DE}}=3 H^2 \left(\frac{2 \alpha-3 \beta}{2+3  \beta  \omega_{_{DE}}}\right). \label{r2}
\end{eqnarray}
The weak energy condition (WEC) leads to $(2 \alpha-3 \beta)
/(2+3\beta \omega_{_{DE}})>0$, imposing constraints on the model parameters due to this inequality. \\
Let us note that the cosmological constant scenario is included as a particular case, given that for $\omega_{_{DE}}=\frac{1}{3\beta}\left( (2\alpha-3\beta)H^2/\Lambda-2\right)$, one can solve (\ref{1Hpt}) and obtain
\begin{eqnarray}
H(t)&=& \sqrt{\Lambda /\alpha} \ \tanh \left( \sqrt{\alpha \Lambda} (t/\beta -c_1) \right) \label{Hl}
\end{eqnarray}
where $c_1=\frac{1}{\beta }-\frac{ 1}{\sqrt{\alpha \Lambda }}$ arctanh $\left(\frac{H_0 \sqrt{\alpha }}{\sqrt{\Lambda }}\right)$, is an integration constant. \\
 Considering the limit at late times when $t$ tends to infinity, from (\ref{Hl}) we obtain $H(t \rightarrow \infty)= \sqrt{\Lambda/\alpha}$ and the dark energy parameter $\omega_{_{DE}}(t \rightarrow \infty)=-1/\alpha$. The associated interaction for this particular case at late times, from the equation (\ref{DE}), is $Q \rightarrow 9 \sqrt{\alpha} (1-\alpha) \Lambda ^{3/2}$ and $r (t\rightarrow \infty)= \frac{2(1-\alpha) }{ 2 \alpha-3 \beta}$, a positive constant.
For the case $\alpha=1$, $\omega_{_{DE}} \rightarrow-1$, $r\rightarrow 0$ and $Q \rightarrow 0$ at late times, thus emulating the $\Lambda$CDM scenario for the late Universe. This particular case that includes $\Lambda$CDM does not solve the cosmic coincidence problem, we will focus on more general scenarios. \\
Considering the result in (\ref{r2}), the cosmic coincidence parameter $r$, can be written as a function of the constants of the model and of the parameter $\omega_{_{DE}}$ as
\begin{eqnarray}
 r=\frac{3H^2}{\rho_{_{DE}}}-1=\frac{2+3 \ \beta \ \omega_{_{DE}}}{ 2 \alpha-3 \beta}-1. \label{r}
\end{eqnarray}
This result allows to establish that giving an Ansatz in the  parameter $\omega_{_{DE}}$ or in $r$ has an equivalence in our formulation.  This is similar to the result found by the authors in \cite{delCampo:2013hka} for non-interacting Ricci holography, \cite{Fei:2013oea} for a given interaction in Ricci-like holography and for interacting MHRDE \cite{Chimento:2011pk}. 
It is noteworthy that this equivalence is a consequence of the holographic nature of the dark energy. The importance of studying variable EoS is relevant in (\ref{r}), since the case for a constant $\omega_{_{DE}}$ would result in a constant $r$ for all time.
The equivalence between $r$ and $\omega _{_{DE}}$ allows simplification of the equation (\ref{1Hpt}) as
\begin{eqnarray}
\dot H= \frac{1}{\beta} \left( \frac{1}{1+r}-\alpha \right)H^2  \label{HH};
\end{eqnarray}
therefore, it is equivalent to consider an Ansatz for $r$, such that (\ref{HH}) can be integrated. \\
By considering the definition of the deceleration parameter $q \equiv -\left( 1+\frac{\dot H}{H^2}\right)$ and the equation (\ref{HH}), we obtain
\begin{eqnarray}
q = \frac{1}{\beta }\left( {\alpha  - \beta  - \frac{1}{1 + r }} \right) , 
\end{eqnarray}
whereas if we have an Ansatz for the coincidence parameter $r(a)$, the deceleration parameter $q(a)$ can be obtained, and thus we can study whether the model is accelerated at late times. It is noteworthy to mention that given that for constant $\omega _{_{DE}}$, we have constant $r$, in this case the desceleration parameter would also be constant, thus not evolving from an non-acclerated Universe to a accelerated one. Note that the parameter $\alpha$ can be rewriten as $\alpha= \beta (q+1)+\frac{1}{1+r}$, value less than one for a positive $r$. \\
Replacing (\ref{r2}) in (\ref{DE}), we obtain an equation to determine the term of interaction,
\begin{eqnarray}
Q = 3H^3  \frac{\left((2-2\alpha+3\beta)r+(3\beta-2\alpha)r^2  + \beta a  r' \right)}{\beta (1+r )^2}, \label{Qr}
\end{eqnarray}
where $\prime$ denotes the derivative with respect to the scale factor. 
The interaction $Q$ has been formulated as a function of $r'$
\cite{Arevalo:2011hh} using $Q=3H \Pi$, where $\prime$ denotes a derivative with respect to the scale factor, and $\Pi$ is a function 
of the energy density $\rho$ and the cosmic coincidence parameter $r$. In this case, the equation was studied exclusively  for $r$ in order to study the late time evolution  of a dynamical autonomous system with a dark energy component, with constant $\omega _{_{DE}}$,  and a dark matter pressureless component. In this work the interaction was positive without giving a function a priori, meaning the constraint $Q>0$ was obtained through the study of the critical points of the system and WEC. Under these conditions it was  obtained that the state parameter is phantom like  ($\omega _{_{DE}}<-1$). These results differ from our case because the parameter $\omega _{_{DE}}$ is not bound to cross the phantom barrier and the sign of $Q$ remains undetermined so far. \\
In the present paper, the result obtained in (\ref{r}) indicates that the evolution of $r$ is determined by the evolution of the Ansatz in $\omega_{_{DE}}$, a relation obtained due to the holographic context in (\ref{HL}). This means that it is equivalent to study the expression for interaction described by (\ref{Qr}) in order to study the interaction as a function of the state parameter $\omega_{_{DE}}$, given as 
\begin{eqnarray}
Q &=&  9H^3 \left(2\alpha-3\beta \right)  \frac{\left( \omega_{_{DE}}   \left(
2\alpha-3\beta-2-3\beta \omega_{_{DE}} \right) +a\beta\omega_{_{DE}}
'\right)}{(2+3 \beta \ \omega_{_{DE}} )^2}. \label{Qw}
\end{eqnarray}
In the expressions (\ref{Qr}) and (\ref{Qw}), it can be observed that the sign of the interaction can change according to the choice of Ansatz ($\omega_{_{DE}}$ o $r$) and the sign of its respective derivatives. \\
There are few number of papers of change of sign for $Q$ in the literature, \cite{Cai:2009ht}, without choosing a special form of interaction $Q$, proposed a new approach in which such interaction is partitioned $Q = 3H\delta(z)$ in the whole redshift range and $\delta(z)$ is constant in each partition. Fitting with observational data they found that the functional $\delta$ probably crosses the non-interacting line ($\delta(z)  = 0$).  The authors suggest that a general interaction term should be studied further in this direction. Following this line of research, the authors \cite{Li:2011ga} proposed a phenomenological interaction of the form $Q(a)=3 b(a)H_0 \rho_0$, such that the coupling function $b(a)$ is variable. With an Ansatz that varies according to the scale factor, the sign change manifests at early times when it is fitted with observational data.
\\
The authors in \cite{Wei:2010cs} study an interaction proportional to the acceleration parameter $q$ and in the work of \cite{Bolotin:2013jpa} they studied some cases of interaction depending of the deceleration parameter and the energy densities. Because of the sign change from a decelerated universe to an accelerated one, a sign change is presented in the interaction. \co{More recently in \cite{Khurshudyan:2013ohq} they consider a generalization of eq. (\ref{pDE}) with a pressure that depends of $(\rho,H,H^2, \dot H)$ and a given interaction that depends on $q$, the energy densities and its derivatives. Considering these interactions $Q=Q(q)$ and comparing it with our work, we find that }there is a relation between interaction and the deceleration parameter; however it is not linear but given as
\begin{eqnarray}
Q  &=& 3{H^3 }\left[ {3\left( {1 - \alpha } \right) - \left( {q  + 1} \right)\left( {2 - 3\beta  - 2\alpha  - 2\beta q } \right)} \right] + \ 3 {H^2} \beta \ \dot q . \label{Qq}
\end{eqnarray}
In our case the sign change depends on the value of $\alpha$ and $\beta$, the acceleration parameter $q$, and its respective derivative $\dot q$. Note that in this case $\beta=0$ would imply a lineal relation between interaction and deceleration parameter. It is the term $\dot H$ that allows a richer dynamics.
Nevertheless, in our case the possibility that the sign changes is obtained as a result of the holographic-like scheme used for the dark energy density (\ref{HL}). If we consider the holographic cut-off for the Hubble-type dark energy $\beta=0$
\cite{Li:2010cj,Li:2004rb}, we obtain
\begin{eqnarray}
Q =-9H^3 \alpha(1-\alpha)\ \omega_{_{DE}} .
\end{eqnarray}
For this case, the interacting function has a defined sign, given by the range $0<\alpha<1$ \footnote{The holographic-type Hubble, $\rho=3H^2 \tilde{c}^2$, has the condition $0<\tilde{c}^2<1$ according to
\cite{Li:2004rb}}, therefore $Q$ is defined positive for all evolution. 
\\
Analyzing in detail the expression (\ref{Qr}) and using the equivalence (\ref{r}), we obtain that the interaction can be simplified as
\begin{eqnarray}
	Q=3H^3 \frac{r}{  (1+r)^2} \left( -3  \omega_{_{DE}}+ a \frac{r'}{r} \right), \label{Qrw}
\end{eqnarray}
where the only negative expression is the term $a\frac{r'}{r}$,  where $r$ is a positive decreasing function that would thus alleviate the cosmic coincidence problem. 
\co{As in the case for \cite{Cai:2009ht} this is a result and not an objective, but in his case it was obtained from observational modelling and not from a theoretical proposal. In addition, for the cases when it was a given sign change interaction,  contrast with late Universe observational data has been done for \cite{Li:2011ga} and \cite{Wei:2010cs}.
}

\section{Ansatzes on $\omega_{_{DE}}$ and observational data}
We consider different Ansatzes for the dark energy parameter $\omega_{_{DE}}$ in order to study in detail the system described by (\ref{1Hpt}). 
 We present three cases which fulfill the criteria that $H$ can be integrated analytically and that such a result respects WEC for some range of the parameters $\alpha$ and
$\beta$. In \cite{Zhang:2013mca}, the authors consider differents cutt-off and impose the energy conditions to constraint the parameter of each holography. We will use this approach in the present work. We summarize three cases in Table 1. 
\begin{table}[!ht]
\begin{center}
\begin{tabular}{ l l } 
\hline\noalign{\smallskip}
   Ansatz & $\omega_{_{DE}} $    \\ 
	\noalign{\smallskip}\hline\noalign{\smallskip}
   Case I & $\omega _0+\eta (1-a)$      \\ 
   Case II & $ \omega _0+\eta (1-a)/a$   \\ 
   Case III & $ \omega_0 +\eta (a^{-3 \epsilon}-1)$    \\ 
	\noalign{\smallskip}\hline\noalign{\smallskip}
 \end{tabular}
 \label{tabla1}
\end{center}
 \caption{The parameters $\omega _0$, $\eta$ and $\epsilon$ are constants to be determined by using the Type Ia supernovae, where
$\omega_{0}$ is $\omega_{_{DE}}$ today ($a=1$)}
\end{table}
\\ We know the cosmic coincidence parameter $r$ from equation (\ref{r}). With this we can evaluate how the interaction function $Q$ behaves according to equation (\ref{Qr}) or equation (\ref{Qw}).
The Hubble parameter $H$ can be obtained and then the parameter space where it is positive defined. In the three cases in Table 1 we know $\rho_{_{DE}}$, $\rho_{_{DM}}$, so we can calculate ranges for the parameters where $H$ is well defined (positive and real). We will study the possible presence of singularities for the solutions of each Ansatz, which will be classified according to the scheme described in \cite{Nojiri:2005sx}. For each case the upper index $I$, $II$ and $III$ is used for the variables of the system $a$ and $\omega_{_{DE}}$.
\\
To estimate the value of the parameters of the model in each case of the Table 1, we will a data set of late universe observations  within the range $(0.1<z<1.4)$. In our adjustment with type I Supernovae \footnote{Analysis with the data set H(z) was performed, but since supernovae has more data points we report these.} called Union 2 \cite{Kowalski:2008ez}, we use Bayesian statistics for $N=4$ free parameters and we marginalize over the constant value of $H_0$, according to the description in \cite{Verde:2007wf}. 
With the Hubble parameter $H$ we construct the theoretical luminosity distance $d_L(theo)$ and we compare it with supernovae luminosity distance through the function
\begin{eqnarray}
\chi    ^2=\sum _i \frac{\left(d_L^{theo}-d_L^{obs} \right)^2}{\sigma^2}
\end{eqnarray}
where $\sigma$ is the error associated with the supernova compilation.
The function $\chi^2$ depends on the parameters of the system and not the redshift $z$. The parameters of the model that best fit the data can be obtained by minimizing the function $\chi^2$, whose minimum value is denoted as $\chi^2_{min}$. Based on this minimum we defined the reduced $\chi^2$ as
\begin{eqnarray}
    \chi^2_{*}=\frac{\chi ^2_{min}}{n-N}, \label{chi}
    \end{eqnarray}
where $n$ is the number of data points of both data sets and $N$ is the number of parameters. A good fit is obtained when the value (\ref{chi}) is close to one ($\chi^2_{*}\leq 1$). \co{Constraints on the parameters of holographic dark energy has been done for several model using observational data by the authors in \cite{Hu:2006ar}, \cite{Zhang:2013mca}, \cite{Gao:2007ep}, \cite{Chimento:2011pk}, \cite{delCampo:2013hka} and \cite{Wang:2010kwa}.}

\subsection{Case I}
In this case we use the parametrization Chevalier-Polarski  $\omega_{_{DE}}=\omega_{0}+\eta(1-a)$
\cite{Chevallier:2000qy,Linder:2002et}, initially introduced to fit Type Ia Supernovae data. Thus the system is determined, then
(\ref{1Hpt}) can be integrated analytically and leads to
\begin{eqnarray}
H(a) &=& H_0 \   a^{-3\frac{(1+\alpha (\eta+\omega_0))}{(2+3 \beta (\eta+\omega_0))}}   \left(\frac{2-3(a-1)\eta \beta+3 \omega_0 \beta}{2+3 \omega_0 \beta} \right)^{\frac{3\beta-2 \alpha}{\beta (2+3 (\eta+\omega_0)\beta )}} \hspace{-0.3cm}, \label{H1}
\end{eqnarray}
where $H_0$ is defined as $H(1)$, thereby redefining the integration constants.
Analyzing the expression obtained for the Hubble parameter in  (\ref{H1}), it can be seen that there is a singularity for $a^{I}_s = \frac{2}{3\beta \eta _0} + \frac{\omega _0}{\eta } + 1$. This is a type three singularity according to the classification \cite{Nojiri:2005sx}, given that for a finite value of the scale factor $a^{I}_s$, the energy density and the pressure diverge.
This singularity will not happen in the future evolution within the validity of our model because there is a certain value $a^{I}_c= \left( 1-2\alpha/(3 \beta) \right)/ \eta+a^{I}_s$, where the coincidence parameter is null $r(a_c)=0$, which shall be proved later on to occur earlier.
\\
From (\ref{H1}) we obtain a range for $\alpha$, $\beta$, $\eta$ and $\omega_0$ as summarized in the following inequalities
\begin{eqnarray}
\frac{\beta }{ \alpha } < \frac{2}{3}   , \    \beta  <   \frac{ -2}{3\omega _0}   , \ \beta  <   {\frac{-2}{3(\omega _0 + \eta)}}  , \ \alpha  <   \frac{-1}{(\omega _0 + \eta )}, \label{wecI}
\end{eqnarray}
where other ranges were rule out, because they did not allow a viable cosmological model. The first inequality of (\ref{wecI}) allow us to infer that
\begin{eqnarray}
a^{I}_c-a^{I}_s=\left( 1-\frac{2}{3}\frac{\alpha }{ \beta} \right)<0
\end{eqnarray}
i.e., the value $a^{I}_c$ is less than the value of the singularity $a^{I}_s$. As a result of this, our model for this Ansatz never reaches the singularity.
\\
We perform the observational analysis with Type Ia supernova data for the parameters $\omega_0$, $\eta$, $\alpha$ and  $\beta$, and the results are presented in Table 2. 
\begin{table}[!ht]
\begin{center}
\begin{tabular}{ l l l } 
\hline\noalign{\smallskip}
    Parameters &  $CI.1$ & $CI.2$ \\ 
	\hline\noalign{\smallskip}
     $\omega_0$    & -1.23&-1.001\\ 
		$\eta $ &0.9 & 0.999 \\ 		
   $\alpha$ &   0.95 & 0.86\\ 
   $\beta $ & 0.24 & 0.23\\ 
    $\chi ^2_* $  & 0.982& 0.982\\ 
\hline\noalign{\smallskip}
 \end{tabular} \\
 \label{tabla2}
\end{center}
\caption{Table summarizing the results for Supernovae analysis for the Case I}
\end{table}
\\ The second condition ($CI.2$) restricts the space parameters to the subspace, where the WEC inequalities (\ref{wecI}) are fulfilled as such that the Hubble parameter is well defined and free of singularities. The dark energy evaluated currently is about $79\%$, compatible with what is observed today. The model presents a singularity at $a^{I}_s=2.9$, but the range of validity of the model is up to $a^{I}_c=1.4$. The countours levels are plotted for the 2-parameter space ($\alpha-\beta$) for the case $CI.2$ (Fig. 1).
\begin{figure}[!ht]
	\centering
	\includegraphics[width=0.4\textwidth]{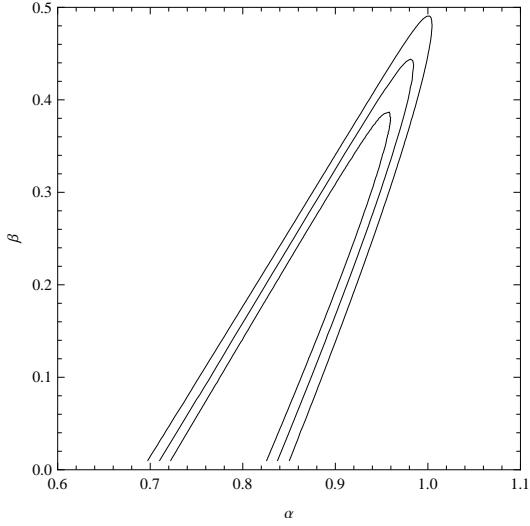}
	\caption{Countours levels are plotted for the 2-parameter space ($\alpha-\beta$) for the case $CI.2$, considering $\omega_0=-1.001$ and $\eta=0.999$}
		\label{CL1}
\end{figure}
\\ The graphic \textit{$Q/(3H^3)$ vs $a$} (Fig. 2) has a crossing at $a^{I}_{_Q}=1.07$, i.e. it changes sign at a definite time. The parameter of deceleration $q$ changes sign when $a^{I}_{q}=0.53$, which indicates a passing from a non-accelerated evolution to an accelerated evolution.
\begin{figure}[!ht]
	\centering
	\includegraphics[width=0.45\textwidth]{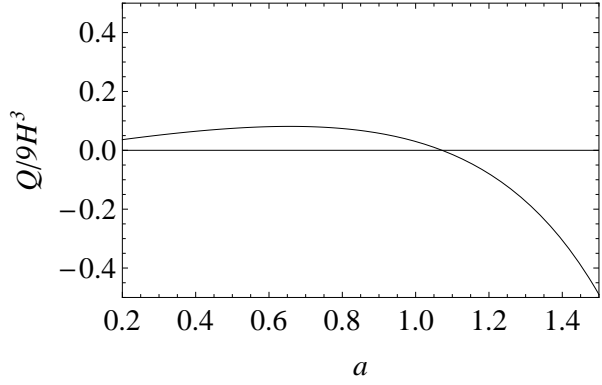}
	\caption{ Interaction Function for the case $CI.2$ versus the scale factor $a$, where one sign change is observed at $a^{I}_{_Q}=1.07$. }
		\label{figura1}
\end{figure}

\subsection{Case II}
The system is determined with the parametrization $ \omega_{_{DE}}=\omega _0+\eta (1-a)/a$ \cite{Virey:2004pr,Virey:2004kh}, which is also used to model Supernova data. Thus, (\ref{1Hpt}) can be integrated and leads to
\begin{eqnarray}
H(a)= H_0 a^{\frac{-\alpha}{\beta }} \hspace{-0.1cm}\left(\frac{(2+3 \omega_0 \beta-3\beta \eta)a+3\beta \eta}{2+3\beta \omega_0} \right)^{\frac{3-2 \alpha/ \beta}{3\beta (\eta-\omega_0)-2 }} \hspace{-0.5cm} . \label{2H}
\end{eqnarray}
Note that the parameter $\beta$ must not be null so the Hubble parameter is well defined. In the same context as in the case I, we analyze the expression obtained in (\ref{2H}). There is a singularity at $a^{II}_s =3 \beta \eta /(2+3 \omega_0 \beta-3\beta \eta)$. This is a type three singularity, because for a finite value of the scale factor $a^{II}_s$, the energy density and the pressure diverge. There is also a certain value $a^{II}_c= 3 \beta \eta /\left( 2\alpha-2-3\beta (1+\omega_0 -\eta)\right)$, where the coincidence parameter is null $r(a_c)=0$. From (\ref{2H}) we obtain the following inequalities
\begin{eqnarray}
	\frac{\alpha}{\beta}>\frac{3}{2} , \quad \beta <\frac{-2}{3 \omega_0} , \quad  \beta >\frac{2}{3 (\omega_0-\eta)}.  \label{wecII}
\end{eqnarray}
We perform the observational analysis on (\ref{2H}) of the parameters $\omega_0$, $\eta$, $\alpha$ and $\beta$. In the table 3 the general result
 is summarized considering all the parameters independent with $\beta>0$ and taking (\ref{wecII}) into account.
\begin{table}[!ht]
\begin{center}
\begin{tabular}{ lll} 
\hline\noalign{\smallskip}
    Parameters & $CII.1$ & $CII.2$\\ 
		\hline\noalign{\smallskip}
     $\omega_0$   & -0,92&  -1.29\\
 $\eta$  & 0.77 & 0.47 \\ 
   $\alpha$  & 0.81 & 0.73\\ 
   $\beta $  & 0.01 & 0.38 \\ 
    $\chi ^2_* $   & 0.9808 & 0.9817 \\ 
\hline\noalign{\smallskip}
 \end{tabular} 
 \label{tabla3}
\end{center}
\caption{Table summarizing the results for Supernovae analysis for Case II}
\end{table}
\\ When WEC is included as a constraint for the parameter space we obtain  the column $CII.2$, with $\chi^2_{*}=0.9817$ and a phantom like component for the EoS parameter $\omega_0 =-1.29$. For this case, the countours levels are plotted in Fig. 3 and the singularity occurs for the value $a^{II}_s=83.72$; therefore, there is a singularity $a^{II}_s$ that cannot be reached in the future because our model is valid until $a^{II}_c=1.64$. 
\begin{figure}[!ht]
	\centering
	\includegraphics[width=0.4\textwidth]{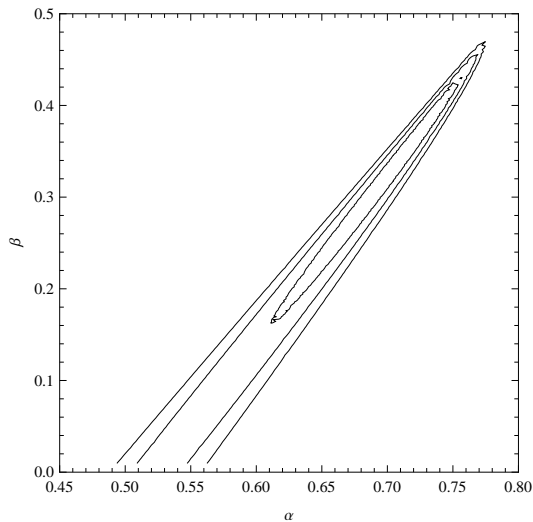}
	\caption{Countours levels are plotted for the 2-parameter space ($\alpha-\beta$) for the case $CII.2$}
		\label{CL2}
\end{figure}
 The interaction function graph presents two sign changes at $a^{II}_{_{Q+}}=0.35$ and $a^{II}_{_{Q-}}=1.25$. That the function $Q$ crosses the non-interacting line twice is introduced by the term $\dot H$ ($\beta \neq 0$) in (\ref{Qq}), which is a equation with two roots for this case.
 The deceleration parameter $q$ changes sign when $a^{II}_q=0.58$, which indicates a passing from an a decelerated evolution into an accelerated evolution.
 
\begin{figure}
	\centering
			\includegraphics[width=0.45\textwidth]{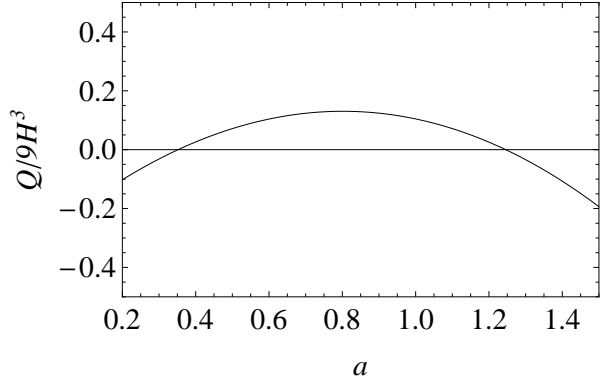}
		\caption{Interaction Function for the case $CII.2$ versus the scale factor $a$, where two sign changes are observed at $a^{II}_{_{Q+}}=0.35$, in the past, and $a^{II}_{_{Q-}}=1.25$, in the future.}
		\label{figura2}
\end{figure} 

\subsection{Case III}
Finally, we consider the coincidence parameter of the form $r=r_0a ^{-3 \epsilon}$ \cite{Zimdahl:2002zb} as an Ansatz, where $\epsilon$ is a positive constant and $r_0 \equiv \rho_{_{DM0}}/\rho_{_{DE0}}$ is the current value for $r$. Considering that there is a relation between $r$ and $\omega_{_{DE}}$ given by the equation (\ref{1Hpt}), this is equivalent to using the Ansatz $\omega_{_{DE}} = \omega_0 +\eta (a^{-3 \epsilon}-1)$.
\\
For simplicity, we use the Ansatz in $r$, where the expression for the Hubble parameter is
\begin{eqnarray}
H(a) &=& H_0 \ a^{\frac{1}{\beta }(1-\alpha)} \left(\frac{r_0 a^{-3 \epsilon} +1}{r_0+1}\right)^{\frac{1}{3\beta \epsilon}}. \label{H3}
\end{eqnarray}
In this case, the Hubble parameter is well defined for all the parameter space, since $r_0$ is by definition a positive parameter. We consider this model free of singularities since $H$ is well defined for all $a$. The expression (\ref{H3}) is the function we compare with observational data, so we will adjust the parameters $\alpha$, $\beta$, $r_0$ y $\epsilon$.\\

\begin{table}[!ht]
\begin{center}
\begin{tabular}{ l l l} 
\hline\noalign{\smallskip}
    Parameters & $CIII.1$ & $CIII.2$  \\
	\hline\noalign{\smallskip}
     $\epsilon$  &  0.030 & 0.107 \\ 
 $r_0$ &    0.728 & 0,365\\ 
   $\alpha$   & 0.572 & 0.714 \\ 
   $\beta $   &  0.018 & 0.050 \\ 
    $\chi ^2_* $   & 0.9813 & 0.9814 \\ 
		\hline\noalign{\smallskip}
 \end{tabular} \\
 \label{tabla4}
 \end{center}
\caption{Table summarizing the results of Supernovae analysis for Case III}
\end{table}

For both cases, $CIII.1$, the general fit and $CIII.2$, for $r_0<0.5$, we obtain $\chi_* \sim 0.981$, where both results fit type Ia supernovae data as shown in Table 4. However, we will work on the graphics with the second result,  because $r_0=0,365$ is the most compatible with other data sets of the universe today \cite{Komatsu:2010fb}. The countours levels of this case are presented in Fig. 5.

\begin{figure}[!ht]
	\centering
	\includegraphics[width=0.4\textwidth]{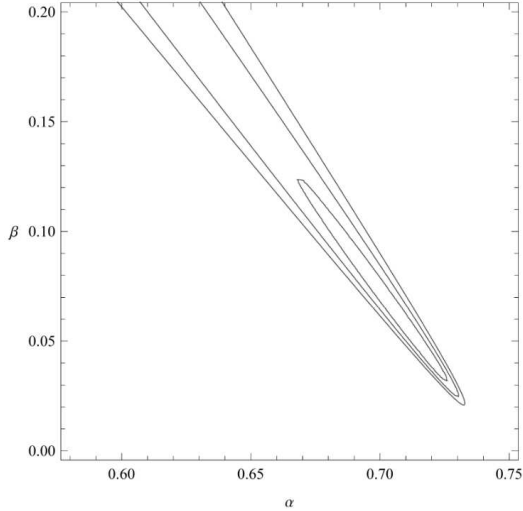}
	\caption{Countours Levels are plotted for the 2-parameter space ($\alpha-\beta$) for the case $CIII.2$}
		\label{CL3}
\end{figure}

The interaction function in Fig.6 presents a change of sign at $a^{III}_{_Q} =0.68$ and displays a similar behaviour to the graphic presented by \cite{Li:2011ga}.  The deceleration parameter $q$ changes sign when $a^{III}_{q}=0.7$, which indicates a passing from a decelerated evolution to an accelerated evolution.

\begin{figure}
	\centering
	\includegraphics[width=0.45\textwidth]{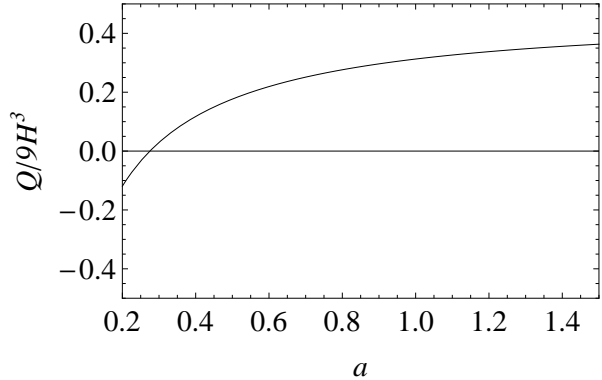}
		\caption{Interacting function for the column $CIII.2$,  where the sign change of the function is at  $a^{III}_{_Q}=0.68$}
		\label{figura3}
\end{figure}

\section{Discussion}
We studied a holography for two interacting fluids in a flat FRW Universe. The interaction $Q$ is not a phenomenological function given a priori, but an unknown variable to be determined according to the nature of the dark energy.\\
In the Ricci-like holographic scheme, there is an equivalence between the coincidence parameter $r$ and the dark energy equation of state $\omega_{_{DE}}$, given by the equation (\ref{r}). 
\\
From this equation it is observed that the cosmic coincidence problem is alleviated, given that when  $\omega_{_{DE}}$ is a variable function, $r$ is also a variable function, and (\ref{r}) can tend asymptotically to a positive constant.
\\
The holographic cut-off Ricci-like ($\beta \neq 0$) induces a change of sign in the interaction function $Q$, illustrated in equation (\ref{Qrw}), which is valid for any Ansatz for $r$ o for $\omega_{_{DE}}$. This phenomenon has not been found yet in other cut-off. In addition, there is a relation between the interaction $Q$ and the deceleration parameter $q$ through the differential equation (\ref{Qq}). This equation shows that $Q$ is a linear combination of ($q$, $q'$, $H$) and the transition of $Q$ imposes conditions on the evolution of the parameter $q$ and its derivative.
\\
This model enables, that for interacting cases, the study of a plethora of possible cases, thus generalizing the result obtained in \cite{delCampo:2013hka}. To parametrize the dark energy we used Ansatz for  $\omega_{_{DE}}$, introduced initially to adjust Supernovae data, in such a way as to make the system integrable. These Ansatzes introduce two additional constants, then each case has four parameters to adjust.
\\
For Case $I$, the best fitting of the parameters was $(\alpha, \beta, \omega_0,\eta)$=$(0.86,0.23,-1.001,0.99)$, with $\chi^2_*=0.982$. Considering this result for the parameters, we evaluate the singularity $a_s^{I}=2.9$ and the cut at $a_c^{I}=1.4$. WEC impose a restriction on the parameters, which is subsequently used as a constraint in the observational analysis. For the interaction $Q$, there is a sign change in the future at $a_{_Q}^{I}=1.07$, from $Q>0$ to $Q<0$, as can be seen in Fig. 2.
\\
For Case $II$, the best fit for the parameters was $(\alpha, \beta, \omega_0,\eta)=(0.73,0.38,-1.29,0.47)$, thus obtaining a value of $\omega_0$ phantom like, with  $\chi^2_*=0.981$. With these values for the  parameters, we evaluated the singularity at $a_s^{II}=83.72$ and the end of the validity of the model at $a_c^{II}=1.64$.  WEC impose a restriction on the parameters, that was later on used as a constraint in the observational analysis. For the interaction $Q$, there are two sign changes, $Q^{II}$, one in the past, for $a^{II}_{_{Q+}}=0.35$ (from $Q<0$ to $Q>0$) and the other is in the future, for $a^{II}_{_{Q-}}=1.25$ (from $Q>0$ to $Q<0$), as shown in the Fig. 4. The interaction decreases until the validity range of the models ends at $a_c^{II}$.
\\
For Case $III$, the best fit is for the parameters $(\alpha, \beta, r_0, \epsilon)=(0.714,0.050,0.365,0.107)$ with  $\chi^2_*=0.981$. Using the Ansatz for $r$ allows a model free of singularities with a reasonable value of $r_0$ for cosmic coincidence. For the interaction $Q$, there is a change of sign in the past, for  $a^{III}_{_Q}=0.68$ , from $Q<0$ to $Q>0$, as can be appreciated in Fig. 6. In the future the interaction $Q$ decreases asymptotically to zero and from early times until today behaves similarly to the graphic of the interaction presented by \cite{Li:2011ga}.
\\
In the cases that present singularity, the moment at which occurs $a_s$ is not included within the range of validity of the model, it occurs in the future ($1<a_c<a_S$) and the interaction evolves in such a way that the transference is from dark matter into dark energy for $a_{_Q} \leq a \leq a_c$. 
\\
Regarding the cases studied, imposing conditions on the parameters in the search for minimums modifies the result and the behavior of the model. This is interpreted as reducing the parameter space into one that has physical validity, then the minimum presented is within such a region. Values of $\alpha$ and $\beta$ are of the same order of magnitude as the fit obtained for the Ricci holography case \cite{Zhang:2013mca}. The parameters associated with the equation of state of the dark energy, $\omega_0$,  $\eta$ and $\epsilon$ are compatible with the current view of the universe. The sign of the function $Q$ in the three cases presented changes at some point in its evolution, \co{it is noteworthy to mention that this is not a given characteristic (as previously mentioned in Section (\ref{dos})), but rather a direct consequence of the holographic context given by a Ricci-like dark energy model.}

\section*{Acknowledgements}
This work has been supported by Comisi\'on Nacional de Ciencias y Tecnolog\'ia
through Fondecyt Grants 1110076 (SL) and 3130736 (FA). \co{(SL) thanks Vicerrector\'ia de Investigaci\'on y Estudios Avanzados, Pontificia Universidad Cat\'olica de Valpara\'iso for its support through Grant 037377/2014}.
 (FP) and (PC) acknowledges DI12-0006 of Direcci\'on de Investigaci\'on y Desarrollo, Universidad de La Frontera.  (PC) acknowledges partial financial support from the Master of Physics Program of Universidad de La Frontera.  (SL) acknowledges the
hospitality of the Physics Department of Universidad de La Frontera where part
of this work was undertaken. One of the authors (FA) acknowledges A. Cid of the  Universidad del B\'io-B\'io for helpful discussions. The authors thank the anonymous referee for helpful suggestions and references, Helen Lowry for reviewing the english and Springer support for latex assistance.

\end{document}